\documentclass{IEEEtran}
\usepackage{subfig}
\usepackage{graphicx}
\usepackage{balance}
\usepackage{algorithm}
\usepackage{algorithmicx}
\usepackage{amsmath}
\usepackage[noend]{algpseudocode}
\usepackage{amsmath,amssymb,amsfonts}

\usepackage{graphicx}
\usepackage{textcomp}
\usepackage{xcolor}
\def\BibTeX{{\rm B\kern-.05em{\sc i\kern-.025em b}\kern-.08em
    T\kern-.1667em\lower.7ex\hbox{E}\kern-.125emX}}
\usepackage{balance}  
\usepackage{booktabs} 

\usepackage{float} 

\usepackage{algorithm,algpseudocode}

\usepackage{amssymb,amsmath,amsthm}
\usepackage{mathtools} 

\begin{document}
\title{Video Monitoring Queries}

\author{\IEEEauthorblockN{Nick Koudas, Raymond Li, Ioannis Xarchakos}\\
\IEEEauthorblockA{\textit{University of Toronto} \\
\{koudas,raymond.li,xarchakos\}@cs.toronto.edu}
}

\maketitle
\begin{abstract}
Recent advances in video processing utilizing deep learning primitives achieved breakthroughs
in fundamental problems in video analysis such as frame classification and object detection enabling
an array of new applications.

In this paper we study the problem of interactive declarative query processing on video streams.
In particular we introduce a set of approximate filters to speed up queries that involve
objects of specific type (e.g., cars, trucks, etc.) on video frames with associated spatial relationships among them (e.g., car left of truck).
The resulting filters are able to assess quickly if the query predicates are true to proceed
with further analysis of the frame or otherwise not consider the frame further avoiding costly object
detection operations.

We propose two classes of filters $IC$ and $OD$, that adapt principles from deep image classification and object detection.
The filters utilize extensible deep neural architectures and are easy to deploy and utilize.
In addition, we propose statistical query processing techniques to process aggregate queries involving
objects with spatial constraints on video streams and demonstrate experimentally the resulting
increased accuracy on the resulting aggregate estimation. 

Combined these techniques constitute a robust set of video monitoring query processing techniques.
We demonstrate that the application of the techniques proposed in conjunction with declarative queries on video streams can
dramatically increase the frame processing rate and speed up query processing by at least two orders of magnitude.
We present the results of a thorough experimental study utilizing benchmark video data sets
at scale demonstrating the performance benefits and the practical relevance of our proposals.
\end{abstract}
\section{Introduction}
In the last few years, Deep Learning (DL) \cite{DBLP:journals/ijcv/RanzatoHL15,DBLP:journals/nature/LeCunBH15} has become a dominant artificial intelligence (AI) technology in industry and academia. Although by no means a panacea for everything related to AI it has managed to revolutionize certain important practical applications such as machine translation, image classification, image understanding, video query answering and video analysis.

Video data abound; as of this writing 300 hours of video are uploaded on Youtube every minute. The abundance of mobile devices enabled video data capture en masse and as a result more video content is produced than can be consumed by humans. This is especially true in surveillance applications. Thus, it is not surprising that a lot of research attention is being devoted to the development of techniques to analyze and understand video data in several communities. The applications that will benefit from advanced techniques to process and understand video content are numerous ranging from video surveillance and video monitoring applications, to news production and autonomous driving. 

Declarative query processing enabled accessible query interfaces to diverse data sources. In a similar token we wish to enable declarative query processing on streaming video sources to express certain types of video {\em monitoring queries}. Recent advances in computer vision utilizing deep learning deliver sophisticated object classification \cite{Krizhevsky:2017:ICD:3098997.3065386,DBLP:journals/corr/SimonyanZ14a,DBLP:conf/cvpr/SzegedyLJSRAEVR15,DBLP:conf/cvpr/HeZRS16} and detection algorithms \cite{DBLP:conf/iccv/Girshick15,DBLP:conf/cvpr/GirshickDDM14,DBLP:journals/pami/RenHG017,DBLP:conf/iccv/HeGDG17,DBLP:conf/cvpr/RedmonDGF16,DBLP:conf/cvpr/RedmonF17,DBLP:journals/corr/abs-1804-02767}. Such algorithms can assess the presence of specific objects in an image, assess their properties (e.g. color, texture), their location relative to the frame coordinates as well as track an object from frame \cite {DBLP:conf/itsc/KrebsDF17} to frame delivering impressive accuracy. Depending on their accuracy, state of the art object detection techniques are far from real time \cite{DBLP:conf/iccv/HeGDG17}. However current technology enables us to extract a {\em schema} from a video by applying video classification/detection algorithms at the frame level. Such a schema would detail at the very minimum, each object present per frame, their class (e.g., car) any associated properties one is extracting from the object (e.g., color), the object coordinates relative to the frame. As such one can express numerous queries of interest over such a schema.

\begin{figure}[th]
\begin{minipage}{.5 \linewidth}
\centering
\includegraphics[height=0.65\linewidth]{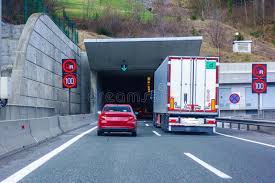}\\
(a)
\end{minipage}%
\begin{minipage}{.5 \linewidth}
\centering
\includegraphics[height=0.65\linewidth]{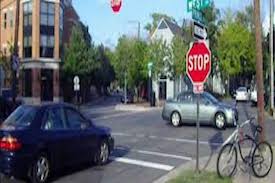}\\
(b)
\end{minipage}
\caption{Example Video Frames: (a) Example Spatial constraints (b) Spatial constraints on a temporal dimension
}
\label{fig:one}
\end{figure}

In this paper we conduct research on declarative query processing over streaming video incorporating specific constraints between detected objects, extending prior work \cite{Kang:2017:NON:3137628.3137664, DBLP:conf/cidr/KangBZ19,blazeit, Hsieh18}. In particular we focus on queries involving {\em count} and {\em spatial constraints} on objects detected in a frame, a topic not addressed in prior art. Considering for example the image in Figure \ref{fig:one}(a) we would like to be able to execute a query of the form \footnote{Adopting query syntax from \cite{Lu:2018:AML:3183713.3183751}}:

\begin{footnotesize}
\begin{verbatim}
SELECT cameraID, frameID, 
C1 (F1 (vehBox1)) AS vehType1,
C1 (F1 (vehbox2)) AS vehType2, 
C2 (F2 (vehBox1)) AS vehColor
FROM (PROCESS inputVideo PRODUCE cameraID, frameID, 
vehBox1, vehBox2 USING VehDetector)
WHERE vehType1 = car AND vehColor = red 
AND vehType2 = truck 
AND (ORDER(vehType1, vehType2) = RIGHT
\end{verbatim}
\end{footnotesize}

that identifies all frames in which a red car has a truck on its right. In the query syntax, $C_i$ are classifiers for different object types (vehicle types, color, etc) and $F_i$ are features extracted from areas of a video frame in which objects are identified (using $vehDetector$ which is an object detection algorithm). Naturally queries may involve more than two objects. Numerous types of spatial constraints exist such as {\em left, right, above, below}, as well as combinations thereof. Categorization of such constraints from the field of spatial databases are readily applicable \cite{Papadias:1995:TRW:223784.223798}. Our interest in not only to capture constraints among objects but also constraints between objects and areas of the visible screen in the same fashion (e.g., bicycle not in bike lane, where bike lane is identified by a rectangle in the screen). We assume that the query runs continuously and reports frames for which the query predicates are true. 

Object detection algorithms have advanced substantially over the last few years \cite{DBLP:conf/iccv/Girshick15,DBLP:conf/cvpr/GirshickDDM14,DBLP:journals/pami/RenHG017,DBLP:conf/iccv/HeGDG17,DBLP:conf/cvpr/RedmonDGF16,DBLP:conf/cvpr/RedmonF17,DBLP:journals/corr/abs-1804-02767}.
From a processing point of view if one could afford to execute state of the art object detection and suitable classification for each frame in real time, answering a query as the one above would be relative easy. We would evaluate the predicates on sets of objects at each frame as dictated by the query aiming to determine whether they satisfy the query predicate. After the objects on a frame have been identified along with their locations and types as well as features, query evaluation would follow by applying well established spatial query processing techniques. In practice, such an approach would be slow as currently state of the art object detectors run at a few frames per second \cite{DBLP:journals/pami/RenHG017}.

As a result we present a series of relatively inexpensive filters, that can determine if a frame is a candidate to qualify in the query answer. As an example, if a frame only contains one object ({\em count filter}) or if there is no red car or truck in the image  or there is no car right of a truck in the frame ({\em class location filter}), it is not a candidate to yield a query answer. We fully process the frame with object detection algorithms only if they pass suitably applied filters. Depending on the selectivity of the filters, one can skip  frames and increase the rate at which streaming video is processed in terms of frames per second. The proposed filters follow state of the art image classification and object detection methodologies and we precisely quantify their accuracy and associated trade-offs. Our main focus in this paper is the introduction and study of the effectiveness of such filters. Placement of such filters into the query plan and related optimizations are an important research direction towards query optimization in the presence of such filters. Such optimizations are the focus of future work.

Armed with the ability to efficiently answer monitoring queries involving spatial constraints, we embed it as a primitive to answer another important class of video monitoring queries, namely video streaming aggregation queries involving spatial constraints. Consider for example Figure \ref{fig:one}(b). It depicts a car at the left of a stop sign. From a surveillance point of view we would like to determine if this event is true for more than say 10 minutes. This may indicate that the car is parked and be flagged as a possible violation of traffic regulations. We introduce Monte Carlo based techniques to efficiently process such aggregation queries involving spatial constraints between objects.

In this paper we focus on single static camera streaming video as this is the case prevalent in video surveillance applications. Moreover since our work concerns filters to conduct estimations regarding objects in video frames and their relationships, we focus on video streams in which objects and their features of interest (e.g. shapes, colors)  are clearly distinguishable on the screen for typical image resolutions. As such the surveillance applications of interest in this study consist of limited number of classes and frames containing small numbers of objects (e.g., multiples of tens of objects as in city intersections, building security, highway segment surveillance etc) but not thousands of objects. Crowd monitoring applications \cite{DBLP:journals/corr/SindagiP17} in which frames may contain multiple hundreds or thousands of objects (sports events, demonstrations, political rallies, etc) are not a focus in this work. Such use cases are equally important but require very different approaches than those we propose herein. They are however important directions for future work.

More specifically in this work we make the following contributions:
\begin{itemize}
\item We present a series of filters that can quickly assess whether a frame should be processed further given video monitoring queries involving count and spatial constraints on objects present in the frame. These include {\em count-filters (CF)} that quickly determine the number of objects in a frame, {\em class-count-filters (CCF)} that quickly determine the number of objects on a specific class in a frame and {\em class-location-filters (CLF)} that predict the spatial location of objects of a specific class in a frame enabling the evaluation of spatial relationships/constraints across objects utilizing such predictions. In each case we evaluate the accuracy performance trade-offs of applying such filters in a query processing setting.
\item We present Monte Carlo techniques to process aggregate queries involving spatial query predicates that effectively reduce the variance of the estimates. We present a generalization of the Monte Carlo based approach to queries involving predicates among multiple objects and demonstrate the performance/accuracy trade-offs of such an approach. 
\item We present the results of a thorough experimental evaluation utilizing real video streams with diverse characteristics and demonstrate the performance benefits of our proposals when deployed in real application scenarios.
\end{itemize}
This paper is organized as follows: Section \ref{sec:filtering} presents our main filter proposals. In section \ref{sec:aggr} we present variance reduction techniques for monitoring aggregates introducing multiple control variates. 
Section \ref{sec:exp} presents our experimental study. Finally section \ref{sec:related} reviews related work and Section \ref{sec:conc} concludes the paper.
\section{Filtering Approaches}
\label{sec:filtering}
In this section we outline our main filtering proposals. We assume video streams with a set {\em frames per second (fps)} rate; we also assume access to each frame individually. Resolution of each image frame is fixed and remains the same throughout the stream. Our objective is to process each
frame fast applying filters and only activate expensive object detection algorithms on a frame when there is high confidence that it will belong to the answer set (i.e., satisfies the query), to make the final decision. We first outline a set of filters which we refer to as {\em Image Classification (IC)} inspired by image classification algorithms \cite{Krizhevsky:2017:ICD:3098997.3065386,DBLP:journals/corr/SimonyanZ14a,DBLP:conf/cvpr/SzegedyLJSRAEVR15,DBLP:conf/cvpr/HeZRS16} and then outline a 
set of filters which we refer to as {\em Object Detection (OD)} inspired by object detection algorithms \cite{DBLP:conf/iccv/Girshick15,DBLP:conf/cvpr/GirshickDDM14,DBLP:journals/pami/RenHG017,DBLP:conf/iccv/HeGDG17,DBLP:conf/cvpr/RedmonDGF16,DBLP:conf/cvpr/RedmonF17,DBLP:journals/corr/abs-1804-02767} \footnote{In this paper we assume familiarity with object classification and object detection approaches, architectures and associated terminology \cite{detectiontutorial}.}.  The set of filters we propose are approximate and as such can yield both false positive and false negatives. We will precisely quantify their accuracy in Section \ref{sec:exp}. From a query execution perspective multiple filters may be applicable on a single frame. In this paper we do not address optimization issues related to filter ordering. A body of work from stream databases is applicable for this, including \cite{Babcock:2003:COS:872757.872789,Lu:2018:AML:3183713.3183751}.
Our implementation of filters is utilizing standard deep network architectures and are implemented as branches in accordance to prior work \cite{Hu2017,DBLP:Zhu2017}. 

\subsection{IC Filters}
\label{sec:ic}
Popular algorithms for image classification \cite{Krizhevsky:2017:ICD:3098997.3065386,DBLP:journals/corr/SimonyanZ14a,DBLP:conf/cvpr/SzegedyLJSRAEVR15,DBLP:conf/cvpr/HeZRS16} train a sequence of convolution filters followed by fully connected layers to derive a probability distribution over the class of objects they are trained upon. 
During training the first layers tend to learn high level object features that progressively become more specialized as we move to deeper network layers \cite{DBLP:books/daglib/0040158}. Recent works \cite{DBLP:conf/wacv/BazzaniBAT16,DBLP:journals/pami/CinbisVS17,DBLP:conf/cvpr/OquabBLS14,DBLP:conf/cvpr/OquabBLS15,DBLP:conf/cvpr/ZhouKLOT16} have demonstrated that concepts from image classification can aid also in object localization (identifying the location of an object in the image). We demonstrate that information collected at convolution layers assists in determining the location of specific object types in an image and detail how this observation can be utilized to deliver object counts as well as assess spatial constraints between objects.
A typical architecture for image classification consists of a number of convolution layers \footnote{Many other architectures are possible}. Before the final output layer it conducts global average pooling on the convolution feature maps and uses them as features for a softmax layer, delivering the final classification. 

\begin{figure}
\includegraphics[width=0.487\textwidth, trim={0cm 14cm 0cm 0cm}, clip]{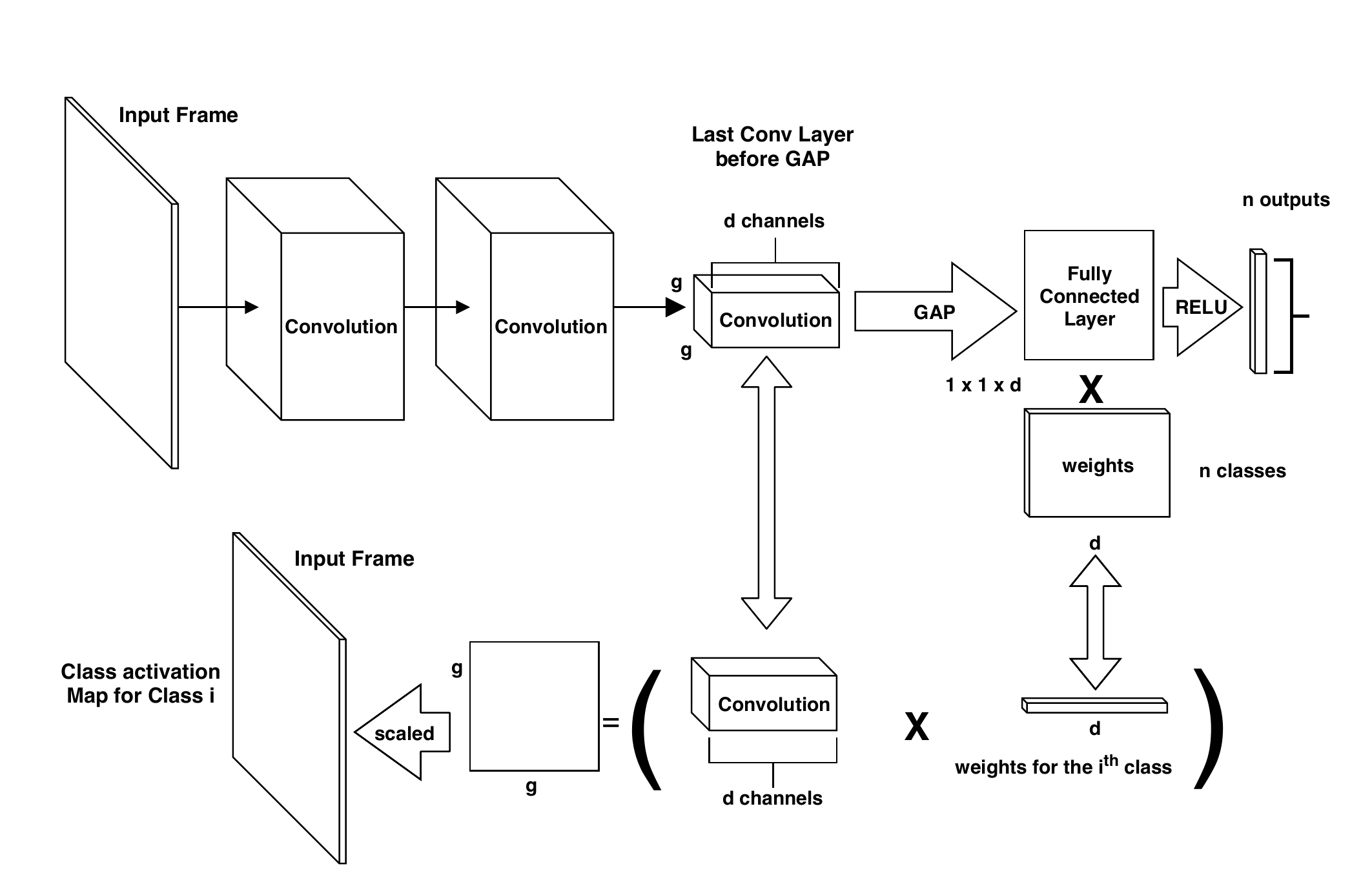}
\caption[Caption for LOF]{\label{fig:oc} Object count prediction architecture. Utilize first few layers of an image classifier (e.g., VGG19). The last feature map $fm$ fees into a fully connected later with ReLU providing object counts per class. The weights of the $i$-th class for the fully connected layer combined with $fm$ provides a $g \times g$ grid for the locations of objects of class $i$.}

\end{figure}

\begin{figure}[th]
\begin{minipage}{.32\linewidth}
\centering
\includegraphics[height=0.55\linewidth]{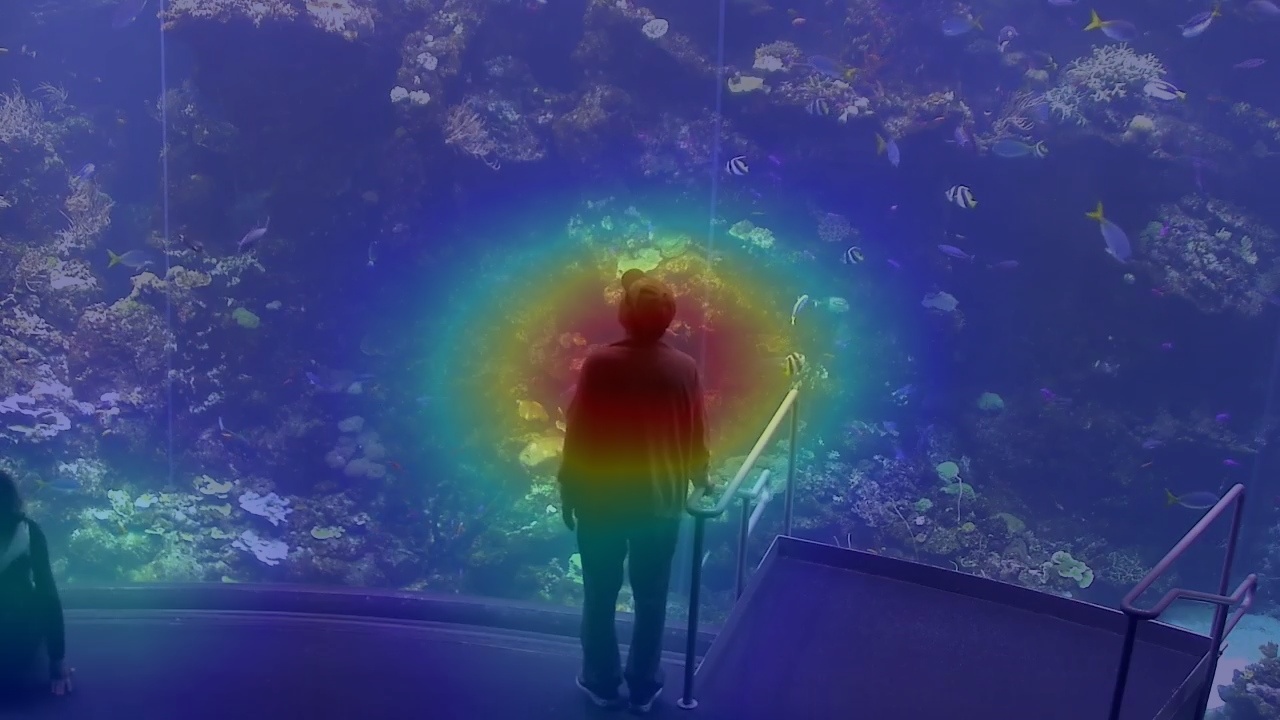}
\end{minipage}%
\begin{minipage}{.32\linewidth}
\centering
\includegraphics[height=0.55\linewidth]{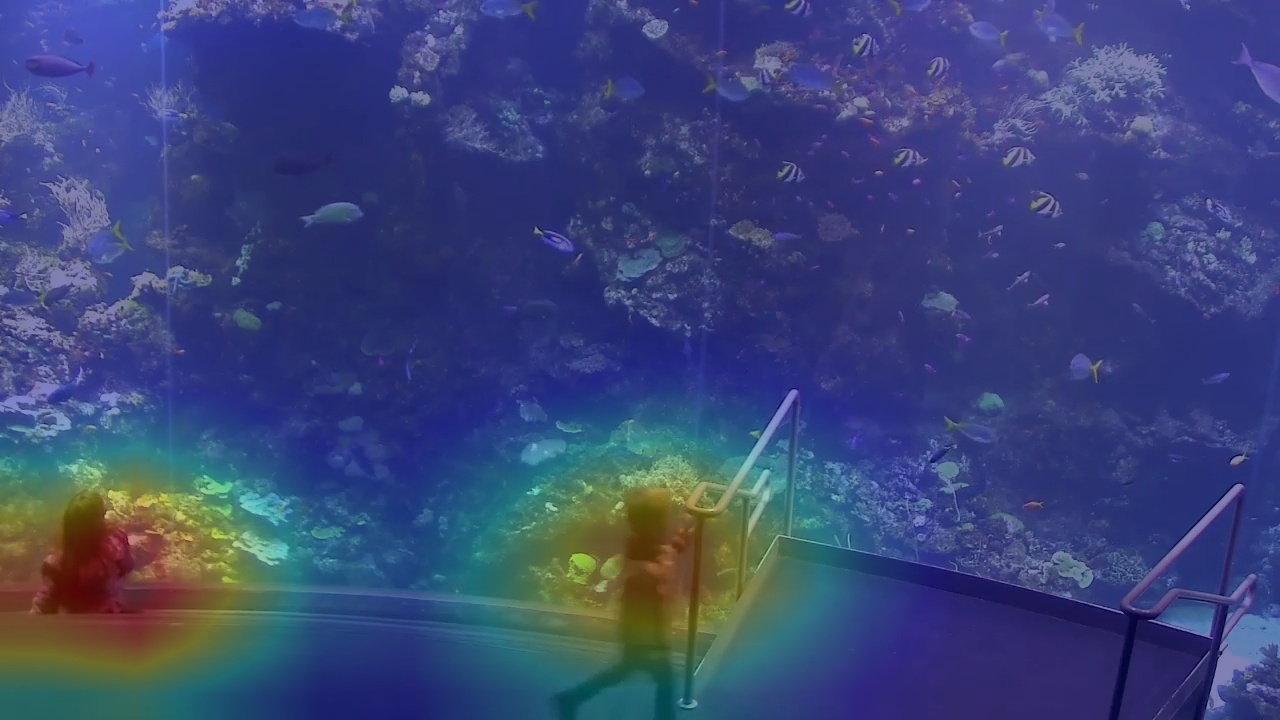}
\end{minipage}
\begin{minipage}{.32\linewidth}
\centering
\includegraphics[height=0.55\linewidth]{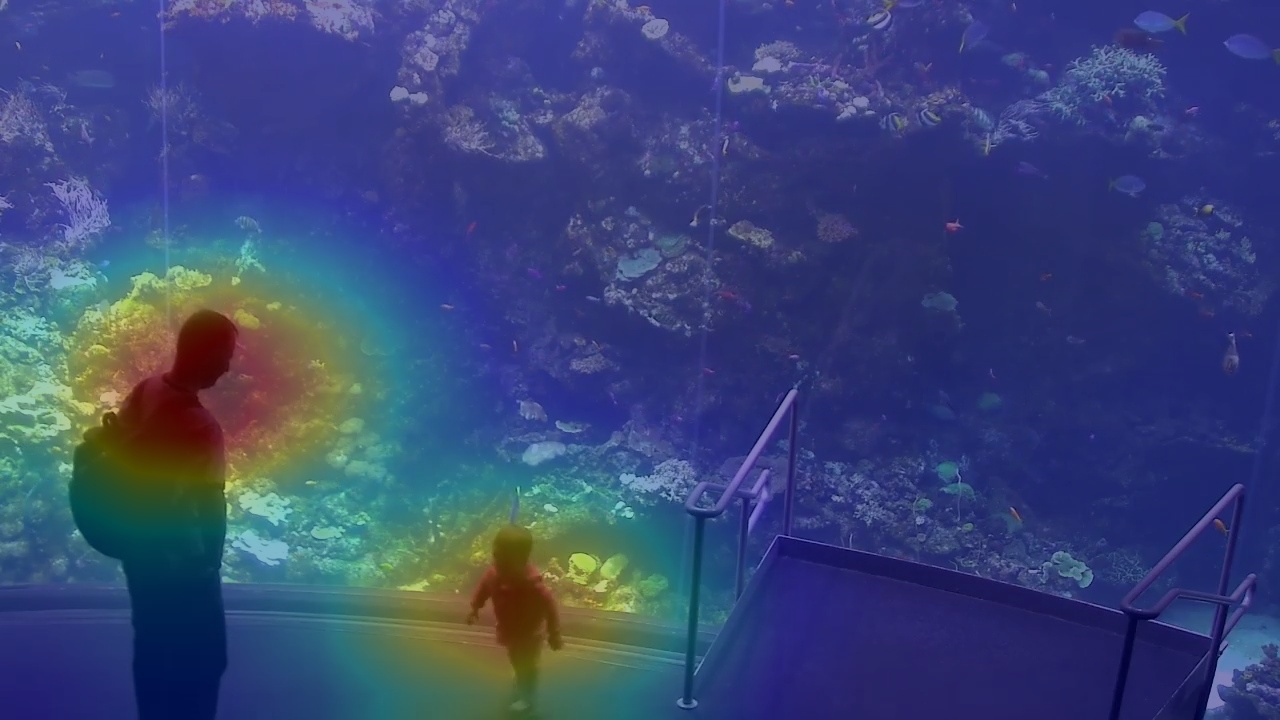}
\end{minipage}
\caption{Sample Class Activation Maps: Localizing Objects exploring neural activation during classification.}
\label{fig:exampleic}
\end{figure}

Assume an image classification architecture (e.g., VGG19 \cite{DBLP:journals/corr/SimonyanZ14a}) with $L=19$ convolution layers and consider the $l-$-th layer from the start.
Let the $l$ convolution layer as in Figure \ref{fig:oc} have $d$ feature maps each of size $g \times g$.
Global average pooling provides the average of the activation for each feature map. Subsequently a weighted sum of these values is used to generate the output. Let $a_k(i,j)$ represent the activation of location $i,j$,$1 \leq i,j \leq g$ at feature map $k$. Conducting global average pooling on map $k$ ($k \in [1 \ldots d]$) we obtain $A_{k} = \sum_{i,j}a_k(i,j)$ (we omit the normalization by $g^2$ to ease notation). Given a class $c$ the softmax input (for object classification) would be $S_c = \sum_k w_k^{c}A_{k}$ where $w_k^{c}$ is the weight for class $c$ for map $k$. As a result $w_k^{c}$ expresses the importance of $A_{k}$ for class $c$. For the class score we have
$S_c = \sum_{k} w_k^{c} \sum_{i,j}a_k(i,j) = \sum_{i,j} \sum_{k} w_{k}^{c} a_{k}(i,j)$. We refer to:
\begin{equation}
\label{eq:1}
M_c(i,j) = \sum_{k} w_{k}^{c}a_{k}(i,j) 
\end{equation}
as the class activation map of $c$ where each element in the map at location $(i,j)$ is $M_c(i,j)$. In consequence $M_c(i,j)$ signifies the importance of location $(i,j)$ of the spatial map in the classification of an image as class $c$. Scaling the spatial map to the size of the image we can identify the image regions most relevant to the image category, namely the regions that prompted the network to identify the specific class $c$ as present in the image. Figure \ref{fig:exampleic} presents an example. It showcases the class activation maps (as a heatmap) for different images
classified as containing a human. It is evident that class activation maps localize objects as they highlight object locations most influential in the classification. We develop this observation further to yield effective filters for our purposes.

Our approach consists of utilizing the class activation map for each object class to count and localize objects. However we improve accuracy
by {\em regulating} the activation map for each class via training. We adopt the first five layers of VGG19 \cite{DBLP:journals/corr/SimonyanZ14a} network pre-trained on ImageNet \cite{Krizhevsky:2017:ICD:3098997.3065386}. 
Experimental results have demonstrated that the first five layers provide a favourable trade off between prediction accuracy and time performance \footnote{Branching at layer 5 provides accuracy close to 90\% requiring around 1ms to evaluate through the first five layers. In contrast placing the branch at layer 15, provides accuracy close to 92\% increasing latency to 1.5ms. Clearly there is a trade-off between accuracy and processing speed which can be exploited as required}.

The fifth layer is a feature map $fm$ of size $d \times g \times g$, where $d=256$ and $g=56$ following the default parameters of the VGG network. We will adopt those, with the understanding that they can be easily adjusted as part of the overall network architecture. In our approach, the feature map is fed through a global average pooling layer followed by a fully connected layer with ReLU activation to output an $n$ dimensional vector that represents the count for each of the $n$ classes. The architecture is as depicted in Figure \ref{fig:oc} (top). In addition the network outputs a class activation map $M_c(i,j)$ of size 56 x 56 for each class $c$  produced by the dot sum product of the feature map $fm$ and the weight vector $w_{k}^{c}$ ($k \in [1 \ldots d]$) (equation \ref{eq:1}) of the fully connected layer, for each $1 \leq c \leq n$. This map is up-scaled to the original image size producing a heat map per class which localize the objects detected for each class. A cell of the map, corresponds to an image area. We threshold the value of each cell of the map in order to detect whether an object of the specific class is present in the corresponding area of the image. That way objects of each class are localized in each cell of the map, allowing us to evaluate spatial constraints between objects as well as between objects belonging to specific classes.

During training we use the following multi-task loss to train the network for performing both count and localization. 
\begin{equation}
\label{eq:lossic}
L=\sum_{\mathrm{c\in{classes}}} weight_c \cdot (\alpha \cdot SmoothL1(x_c, \hat{x_c}) + \beta \cdot MSE(y_c - \hat{y_c}))
\end{equation}

Where $weight_c$ is a weight for class $c$ (computed as the fraction of frames in the training set containing class $c$), $\alpha$ and $\beta$ are parameters for balancing the loss between the two tasks and $x, \hat{x}, y, \hat{y}$ are respectively the count prediction, the count label (ground truth), the class activation map and ground-truth object location map. Notice that involving $\hat{y}$ in the training process {\em regulates} the network to understand object location as well. The training objective utilizes $SmoothL1$ \cite{DBLP:conf/iccv/Girshick15}. The labels for both count and object location maps (ground truth) are generated by Mask R-CNN. Specifically, the location map is produced by down-scaling the locations of the Mask R-CNN bounding boxes in the image to size $56\times{56}$ for each class.

When optimizing for localization, we fix the weights of the fully connected layer and only back-propagated the error to the feature layers of VGG. During training, we first only optimize count by setting parameter $\beta$ to zero, and after 5 epochs, set $(\alpha, \beta)$ to $(1, 10)$, and gradually decrease $\beta$ while keeping $\alpha$ fixed at 1. We found that using this method, the network converges much faster than optimizing both tasks from the start. The training process is optimized with Adam Optimizer \cite{DBLP:journals/corr/KingmaB14} with the only data transformation being normalization and random horizontal flip. 

Obtaining estimates for our various filters follows naturally from the network output. The $IC-CF$ (count of all objects) as well as $IC-CCF$ (class count filter. i.e., count of objects per class) is obtained from the output of the network as it provides a vector with the count of the
objects per class detected in the frame. In addition  $IC-CLF$ (class location filter) are obtained by manipulating the class activation maps for each object class directly. An activation map for class $c$ after thresholding is a binary vector of size $g \times g$ ($56 \times 56$ in our case) indicating whether or not there is an object of class $c$ in the corresponding area of the image. Spatial constraints between objects can be evaluated in a straightforward manner manipulating the thresholded activation maps.

\subsection{OD Filters}
\label{sec:od}
State of the art object detection algorithms typically work as follows: they apply a series of convolution layers (each with a specific size and number of feature maps) to extract suitable features from an image frame, followed by the application of a sequence of region proposals (boxes of certain width and height) to determine the exact regions inside an image that contain objects of a specific class. Although passing an image through convolution layers for feature extraction is relatively fast, the later steps of assessing object regions and object class are where the bulk of the computation is spend \cite{DBLP:conf/iccv/HeGDG17}.

We observe that from an estimation point of view however, being able to assess object count as well as spatial constraints among objects, does not necessarily require precise identification of the entire objects. $IC$ filters utilized object location only via {\em regulation} during training. There is no other object location information on the network as object classification networks are not designed to deal with object detection. We thus explore a hybrid approach. We utilize the solid capabilities of object detection techniques to localize objects of specific type in a frame, without the overhead to discover the object extend in the image. Discovery of the extend is the most time consuming part of object detection \cite{DBLP:journals/cviu/GirshickKLMPVWY17} and not required for estimation purposes. Thus, we propose to utilize an object detection network augmented with a branch to conduct object localization and class detection utilizing only the first few convolution layers.

\begin{figure}
\includegraphics[height=6cm,width=0.5\textwidth, trim={0cm 0cm 0cm 0cm}, clip]{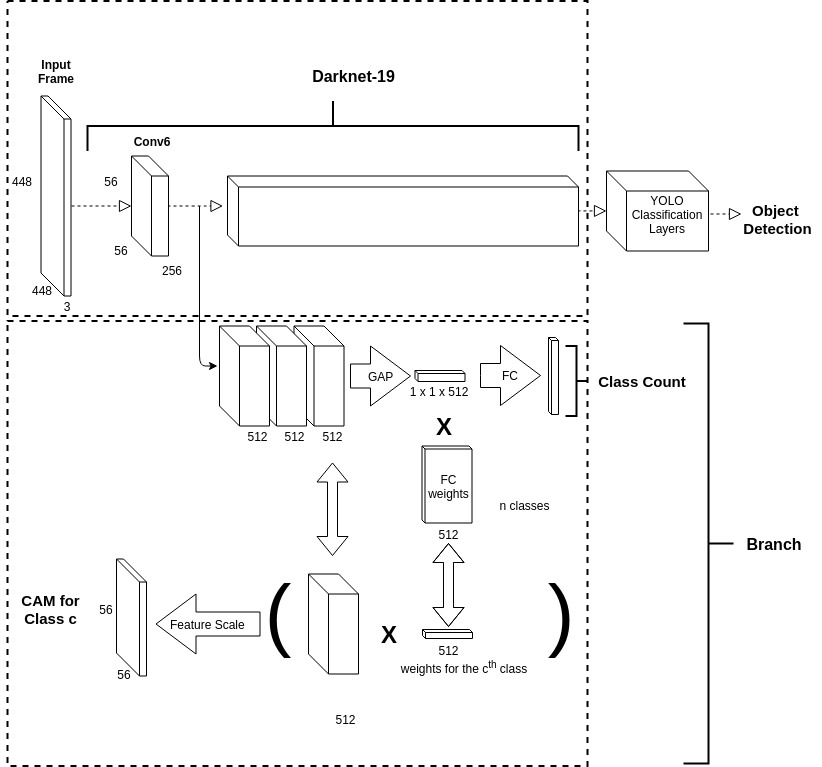}
\caption{\label{fig:od} Object Center prediction architecture. Utilize first few (e.g.,six) layers of an object detector (e.g., YOLOv2) and branch out to the proposed network (with $g$=56) utilizing the features for object prediction on a grid. Branching can take place at later layers as well. }
\end{figure}

The approach is summarized in figure \ref{fig:od}. Our basic architecture consists of a few convolution layers from an object detection framework. We adopt YOLOv2 \cite{DBLP:conf/cvpr/RedmonDGF16,DBLP:conf/cvpr/RedmonF17,DBLP:journals/corr/abs-1804-02767} but any other framework will be suitable as well \cite{DBLP:conf/iccv/Girshick15,DBLP:conf/cvpr/GirshickDDM14,DBLP:journals/pami/RenHG017,DBLP:conf/iccv/HeGDG17}. YOLOv2 utilizes Darknet-19 with 19 convolution layers. Our approach is to pass an image only through $k$ of such layers extracting features that assist in the estimation of object locations in an adjustable image grid of size $g \times g$. The features extracted after the image passes through $k$ layers are utilized as input into a network created as a branch from the underlying object detection architecture. 
 
The parameters of the network at the branch are depicted in Figure \ref{fig:od}. The network has 
three convolution layers followed by global average pooling and a fully connected layer that produces the final answer. The intuition for this architecture is to utilize the convolution layers learned for detection of objects from the object detection network (YOLOv2 in this case) but instead of proceeding with the exact detection of object extents, branch to a network that uses the regularized heat-map of the $IC$ approach to pin point image frame grid cells that contain objects. We
demonstrate that this improves the detection and localization process, as the network is trained to recognize as well as localize objects from scratch.

We set the image frame input of YOLO to $448 \times 448$ pixels. Thus, if we set the branch at layer eight ($k$=8) its  feature map  is a $256 \times 56 \times 56$ vector. The branch network therefore produces a per-class regression output for count and a $56 \times 56$ ($g$=56) grid. The three convolution layers of the branch network are of size $56 \times 56 \times 512$. The first $k$ layers of Darknet are thus shared between the task of YOLO detection and count/object location estimation.

Prior to training, we use a state of the art object detection network (Mask R-CNN \cite{DBLP:journals/pami/RenHG017}) to annotate the training set and asses ground truth. The object detection network is initialized with pre-trained weights from MS-COCO \cite{DBLP:journals/corr/LinMBHPRDZ14}. The network is trained end-to-end minimizing the sum of the loss function of YOLOv2 \cite{DBLP:conf/cvpr/RedmonF17} and the loss of the branch network
which is defined as:

\begin{footnotesize}
\begin{multline}
L_{\textrm{branch}} = \sum_{c=0}^{\overline{C}} [\lambda_{count} \cdot SmoothL1(\textrm{count}^c, \hat{\textrm{count}^c}) + \\
\lambda_{grid} \cdot \frac{1}{g^2} (\mathbb{A}_{ic}^{obj} \sum_{i=0}^{g^2} \lambda_{obj} \cdot (x_i^c - \hat{x}_i^c)^2 + \\ \mathbb{A}_{ic}^{noobj} \sum_{i=0}^{g^2}\lambda_{noobj} \cdot (x_i^c - \hat{x}_i^c)^2)]
\end{multline}
\end{footnotesize}
In the equation, $count^c$ is the count prediction for the $c^{th}$ class, $x_i^{c}$ is the prediction for an object of class $c$ at the $i^{th}$ cell. $A_{ic}^{obj}$ and $A_{ic}^{noobj}$ are respectively the mask indicating whether or not an object of class $c$ exists at grid location $i$. We sum over a subset of classes $\overline{C}$ from MS COCO, and balance the loss for the counts and the loss from the grid with $\lambda_{count}$ and $\lambda_{grid}$. In addition, we set $\lambda_{noobj}$ and $\lambda_{obj}$ for balancing the false positive and false negative loss when summing over the $g^2$ spatial locations of the grid.

For the case of YOLOv2 we use the training data as annotated by Mask R-CNN when computing the loss (that includes both object types and object extents). For the case of the branch network, the labels for both count and object location maps (ground truth) are generated by Mask R-CNN as well. Specifically, as in the case of $IC$ filters, the location map is produced by down-scaling the locations of the Mask R-CNN bounding boxes in the image to the size of the grid, for each class.

Since the output of the branch network is exactly the same as in the $IC$ approach, the estimates for the $OD$ filters are produced in exactly the same way by manipulating the output of the branch network which provides a vector with the counts of objects per class (for the case of $OD-CF$ and $OD-CCF$ filters). Similarly $OD-CLF$ are obtained in the same way manipulating the activation maps of each class.

\subsubsection{Object Count Classifier}

\begin{figure}
\includegraphics[width=0.5\textwidth, trim={4cm 20cm 5cm 5cm}, clip]{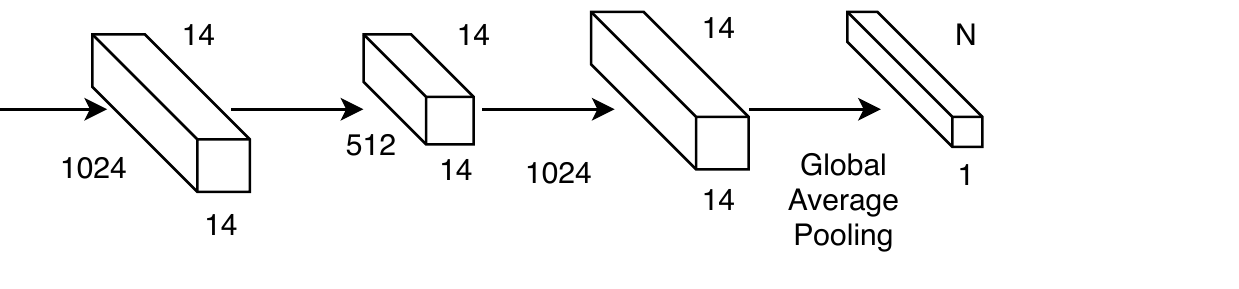}
\caption{\label{fig:odccf} Object Counting Network Branch. Utilize first few (e.g.,five) layers of an object detector (e.g., YOLOv2) and branches out to this network utilizing the features for object count prediction. Branching can take place in later layers as well. }
\end{figure}

We present an alternative approach in which the optimization objective is strictly to predict the number of objects in the image frame. We follow the same methodology as in Section \ref{sec:od} placing a network as a branch of an object detection architecture. The network utilizes the object detection features learned by the $k$-th convolution layer as input. The features are max-pooled to ($F$, $f$, $f$) where $F$ is the number of filters in the $k$-th convolution layer and $f \times f$ the filter size. However the network is exclusively trained to predict the number of objects in the frame and its architecture is depicted in Figure \ref{fig:odccf} and Table \ref{table:2}. We refer to this filter as {\em Object detection, count optimized classification filter} $OD-COF$. 

Prior to training, we are using Mask R-CNN to annotate the training set. To produce the labels for training our model, we obtain the number of objects for each frame detecting all objects and counting them. The proposed network is trained end to end minimizing the loss function. The specifics of training and network parameters are detailed further in Section \ref{sec:exp}. 
\begin{footnotesize}
\begin{table}
\begin{center}
\begin{tabular}{||c|c|c|c|c||}
\hline
Layer & Filters & Size & Padding & Activation \\ \hline
Conv1 & 1024 & 1 & 1 & LeakyReLU \\ \hline
Conv2 & 512 & 3 & 1 & LeakyReLU \\ \hline
Conv3 & 1024 & 1 & 0 & LeakyReLU \\ \hline
Conv4 & 1024 & 1 & 3 & LeakyReLU \\ \hline
\end{tabular}
\caption{\label{table:2} Network parameters for object count prediction}
\end{center}
\end{table}
\end{footnotesize}

\section{Monitoring Aggregates}
\label{sec:aggr}

Real time video streaming, like any streaming data source \cite{Koudas:2003:DSQ:1315451.1315572}, provides opportunities for useful queries when one accounts for the temporal dimension. Object presence along with spatial constraints offer numerous avenues to express temporal aggregate queries of interest. Revisiting the example of Figure \ref{fig:one}(b) one would wish to evaluate the following query \footnote{The query is simplified to ease notation. One has also to account for the {\tt trackid} assigned via object tracking \cite{DBLP:conf/iccv/WangOWL15} to each blue car identified as it enters and leaves the screen so to associate the aggregate to the same blue car in the batch window.}:

\begin{footnotesize}
\begin{verbatim}
SELECT cameraID, count(frameID),
C1(F1(vehBox1)) AS vehType1,
C3(F3(SignBox1)) AS SignType2, 
C2(F2(vehBox1)) AS vehColor
FROM (PROCESS inputVideo 
PRODUCE cameraID, frameID, vehBox1 
USING VehDetector, SignBox1 USING SignDetector)
WHERE vehType1 = car AND vehColor = blue 
AND vehType2 = Stop-Sign 
AND (ORDER(vehType1,SignBox1)=RIGHT)
WINDOW HOPING (SIZE 5000, ADVANCE BY 5000)
\end{verbatim}
\end{footnotesize}

seeking to report the number of frames in a batch window of 5000 frames for which a blue car has a stop sign on its right in a road surveillance scenario\footnote{At typical 30 frames per second (fps) one can deduce when the count is higher than a threshold whether the car maybe parked.}. Similarly one could seek to report the average number of bicycles in a road segment (e.g. bike lane, identified by a bounding box in the frame) per hour. From a query evaluation perspective, one can utilize a sampling based approach evaluating each sampled frame for all objects present (applying a state of the art object detection algorithm) and determining their spatial constraints to identify whether they satisfy the query constraints. Essentially for each frame sampled, we determine a value (boolean or otherwise) that can be utilized in conjunction with sampling based approximate query processing techniques \cite{Chaudhuri:2017:AQP:3035918.3056097,Park:2018:VUA:3183713.3196905,Rusu:2006:FRR:1142473.1142496,Chaudhuri:2007:OSS:1242524.1242526} to derive statistical estimates of interest (e.g., mean, variance estimates as well as obtain confidence intervals for the query answer). 

The various filters introduced thus far, can also be utilized to improve query estimates and in particular reduce the sample standard deviation of the required estimates. We discuss a well known Monte Carlo technique called {\em control variates (CV)} \cite{glasserman} and its application to monitoring aggregate queries. Moreover, since our problem involves queries encompassing multiple objects in a frame and their constraints, we demonstrate how $CV$ can be extended in this case as well introducing {\em Multiple Control Variates}.

Let $Y$ be a random variable whose mean is to be estimated via simulation and $X$ a random variable with an estimated mean $\mu_{X}$. In our trial the outcome $X_i$ is calculated along with $Y_i$. Further assume that the pairs $(X_i,Y_i)$ are i.i.d, then the $CV$ estimator $\hat{Y_{CV}}$ of $E[Y]$ is $\hat{Y_{CV}} = \frac{1}{n} \sum_{i=1}^{n}(Y_i-X_i+\mu_{X})$. It is well known \cite{glasserman} that the $CV$ estimator is unbiased and consistent. The variance of the estimate is $Var(\hat{Y_{CV}}) = \frac{1}{n}Var(Y-X+\mu_{X}) = \frac{1}{n}(\sigma_{Y}^{2} + \sigma_{X}^{2} - 2\rho_{XY}\sigma_{X}\sigma_{Y})$, indicating that the $CV$ estimate will have lower variance when $\sigma_{X}^{2} < 2\rho_{XY}\sigma_{X}\sigma_{Y}$, where $\rho_{XY}$ is the correlation coefficient between $Y$ and $X$. 

To get full advantage of the $CV$ estimate, a parameter $\beta$ is introduced and optimized to minimize $\hat{Y_{CV}}$, namely $\hat{Y_{CV}} = \hat{Y}-\beta({\hat{X}}-\mu_{X})$ with variance $Var(\hat{Y_{CV}}) = \frac{1}{n}(\sigma_{Y}^{2}+\beta^{2}\sigma_{X}^2-2\beta\rho_{XY}\sigma_X\sigma_Y)$. Minimizing for $\beta$ we get $\beta^{*}=\frac{\sigma_{Y}}{\sigma_{X}}\rho_{XY} = \frac{Cov(Y,X)}{Var(X)}$. This reveals that the reduction of variance depends on the correlation of the estimated variable ($Y$) and the control variate ($X$). Substituting $\beta^{*}$ in the expression of variance $Var(\hat{Y_{CV}(\beta^{*}})) = (1-\rho_{XY})\sigma_{Y}^2$.

In practice $Cov(X,Y)$ is never known. However we can certainly estimate sample values from $S_{XX} = \frac{1}{n-1}\sum_{i=1}^{n}(X_i-\hat{X} )^2$ and $S_{YX} = \frac{1}{n-1}\sum_{i=1}^{n}(X_i-\hat{X})(Y_i-\hat{Y})$, providing an estimator $\beta^{*} = S_{XY}S_{XX}^{-1}$.

$CV$'s blend in our application scenario naturally. $Y$ is the variable we like to estimate and $X$ is the result of the application of one of the introduced filters. Since $X$ estimates $Y$ we expect (provided the the filters are good estimators) that the two variables would be highly correlated and as a result $CV$ estimation would provide good results. 
For example, for the case of estimating the average number of bicycles in a bike lane (where the bike lane is identified by a region on the screen) over a time period, $Y_i$ is the result of the application of full object detection for objects falling inside the bike lane region on a frame and $X_i$ is the application of a $CLF$ filter on the frame. The two quantities are estimated by sampling a number of frames over the specified time period (e.g., one hour) applying standard sampling based techniques. In the estimation, we use as $\mu_X$ the sample mean $\hat{\mu_X}$ over the sampled $X_i$'s. The estimated variance however will be tighter utilizing the $CV$ estimates. Thus $CV$ provides a way to trade improved accuracy for small increase in processing time per sample, as in addition to applying the full object detection algorithm we also have to apply the filter. Filter time per sample is a tiny fraction of the object detection time (see Section \ref{sec:exp}), making $CV$ appealing.
Similarly, in the case of {\em car next to a stop sign for more than 10 minutes} query, $Y_i$ is the answer of the application of full object detection followed by processing the identified objects to assess whether there is a car next to a stop sign in the frame and $X_i$ is a $CLF$ that estimates the locations of cars and stop signs on a grid and assess the constraint. Given the frame per second rate of the video, we can identify how many frames constitute the time range of interest (e.g., 10 minutes); then use sampling to identify whether the predicate is true for the entire time interval and the associated variance applying $CV$.

\subsection{Multiple Control Variates}

In certain applications, aggregate monitoring queries can obtain reduced variance estimates by involving multiple control variates. For example consider a query inquiring for the average number of bicycles and trucks entering simultaneously from the right and left of the image respectively in a traffic monitoring scenario. We demonstrate how it is possible to generalize control variates to more than one variables. 

Let ${\bf Z} = (Z_1 \ldots Z_d)$ be a vector of control variates with estimated expected values ${\bf \mu_Z} = (\mu_1 \ldots \mu_d)$. In this case, $\hat{Y_{CV}(\beta)} = \hat{Y}-\beta({\bf Z}-\mu_{Z})$ is an unbiased estimator of $\mu_Y$ \cite{glasserman}. Following the case for a single control variate, the optimal value for $\beta$ if provided by $\beta^{*}=\sum_{ZZ}^{-1}\sum_{YZ}$, where $\sum_{ZZ}$ and $\sum_{YZ}$ are the covariance matrix of {\bf Z} and the vector of covariances between $(Y,{\bf Z})$. In practice the covariance matrices will not be known in advance. Thus they are estimated using samples, $S_{Z^jZ^k} = \frac{1}{n-1}\sum_{i=1}^{n}(Z_i^j-\hat{Z^j})(Z_i^k-\hat{Z^k})$ for $j,k \in [1 \ldots d]$ and $S_{YZ^j} = \frac{1}{n-1}\sum_{i=1}^{n}(Y_i-\hat{Y})(Z_i^j-\hat{Z^j})$, with $j \in [1 \ldots d]$. Following the same methodology as in the case of a single variate one can show that $Var(\hat{Y_{CV}(\beta^{*})}) = (1-R^2)Var(\hat{Y})$ where $R^2 = \frac{\sum_{YZ}^{'}\sum_{ZZ}^{-1}\sum_{YZ}}{\sigma_{Y}^2}$ is the squared multiple correlation coefficient, which is a measure of variance of $\hat{Y}$ explained by ${\bf Z}$ as in regression analysis.

\begin{figure}
\includegraphics[width=0.5\textwidth, trim={2cm 13cm 2cm 9cm}, clip]{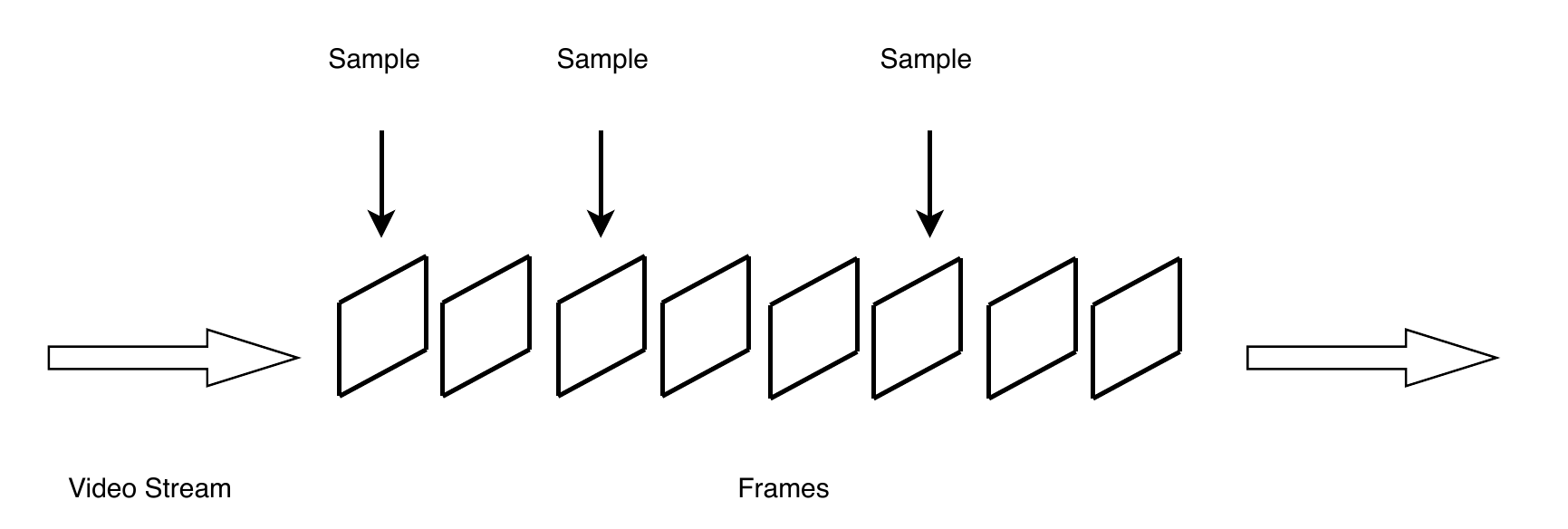}
\caption{\label{fig:aggrexample} A number of frames will be sampled, for each frame suitable all suitable filters are applied and control variates is deployed to estimate the aggregate.}
\end{figure}

For example consider a query seeking to estimate the number of frames in which bicycles and trucks exist having different spatial constraints each (e.g., objects exist in specific screen areas) in frames having more than three cars. In this case we would employ one $CLF$ to obtain an estimate for bicycles, an additional for trucks and one more for car counts and utilize multiple control variates to obtain a reduced variance estimate for this query. Figure \ref{fig:aggrexample} presents the overall approach. In each sampled frame, all applicable filters are applied to qualify the frame. In this case $CLF$ for bicycles and trucks to assess the constraint as well as a count filter for cars. If the frame is qualified by the applicable filter, control variates is utilized for the aggregate.

In section \ref{sec:exp} we will evaluate the ability of $CV$ to reduce variance for different types of aggregate monitoring queries.

\section{Experimental Evaluation}
\label{sec:exp}

We now present the results of a detailed experimental evaluation of the proposed approaches. 
We utilize three different data sets namely, \textbf{Coral}, \textbf{Jackson} (Jackson town square), and \textbf{Detrac}\cite{DBLP:journals/corr/WenDCLCQLYL15}. Of these data sets, \textbf{Coral}, \textbf{Jackson} are video sequences shot from a single fixed-angle camera, while \textbf{Detrac} is a traffic data set composed of 100 distinct fixed angle video sequences shot at different locations. In order to maintain the consistency of our models, we annotate the three data sets using the Mask R-CNN Detector.

The \textbf{Coral} data set is an 80 hour fixed-angle video sequence shot at an aquarium. Similarly, \textbf{Jackson} is a 60 hour fixed-angle sequence shot at a zoomed-in traffic intersection. Finally, \textbf{Detrac} consists of 10 hours of fixed-angle traffic videos shot at various locations in China. We did not use the provided training set label, and instead annotated the data set with various classes of vehicles. 

To evaluate our model, we partition the video sequences to create a training, validation, and test set for each data set. For \textbf{Coral} and \textbf{Jackson}, we create our data set with the mp4 files published by NoScope \cite{Kang:2017:NON:3137628.3137664}. 
The \textbf{Detrac} data set contains 60 and 40 different sequences for training and testing respectively. 
For the purpose of our experiment, we combine the train and test set of the original data set and partition the ordered frames into train, validation, and test set with equal ratios between sequences. Table \ref{table:3} presents a description and key characteristics of each data set.

\begin{footnotesize}
\begin{table}
\footnotesize
\begin{center}
\begin{tabular}{||c|c|c|c|c|l||}
\hline
Dataset & Train Size & Test Size & Obj/Frame & std & Classes \\ \hline
Coral & 52000 & 7215 & 8.7 & 5.1 & Person \\ \hline
Jackson & 14094 & 3000 & 1.2 & 0.5 & \begin{tabular}{@{}c@{}}car (80\%)  \\ Person (20\%) \end{tabular} \\ \hline
Detrac & 55020 & 9971 & 15.8 & 9.8 & \begin{tabular}{@{}c@{}}car (92\%)\\ bus (6\%)\\ Truck (2\%)\end{tabular} \\ \hline
\end{tabular}
\caption{\label{table:3} Datasets and their characteristics}
\end{center}
\end{table}
\end{footnotesize}
With our experiments we seek to quantify the accuracy of each of the filters introduced as well as their effectiveness when applied in conjunction with declarative queries on video streams. We also quantify the variance reduction resulting from the application of the proposed control variates approach on aggregate queries involving various types of spatial predicates. Finally we demonstrate the effectiveness of our approach to identify unexpected objects on video streams, demonstrating image frames flagged as unexpected in sample videos.

For the case of $IC$ filters we utilize VGG19. Placing the branch at layer five results in a grid of size $56 \times 56$. In this case the time required to pass the frame through the network and through the branch to derive our estimations is approximately 1.5ms/frame. Similarly for $OD$ filters we utilize YOLOv2. Placing the branch at layer eight, results in a grid size of $56 \times 56$ as well. The time required to pass a frame through the first eight layers and also through the filter is 1.9ms/frame. For our experiments we choose to place the branches at layers as specified above in order to compare both techniques on at $56 \times 56$ grid. Placing the branch at higher layers introduces a trade-off. Processing
time per frame increases (as we have to pass through more layers). At the same time count accuracy increases (by a few percentage points
according to our experiments) for all techniques as higher level features are discovered in later layers that improve counts. However the grid size becomes smaller due to the network architecture (e.g., $28 \times 28$ or $14 \times 14$ depending on the layer) that penalizes location accuracy (up to 8\% lower across all techniques than the results reported below according to our experiments). We note however that the results do not change qualitatively than those reported below for a grid of size $56 \times 56$.
In comparison the time required per frame by Mask R-CNN is 200ms. Similarly for comparison, passing a frame through the entire YOLOv2 requires 15ms. This provides good localization accuracy for objects (approximately 3-5\% higher than those we report, across data sets for the case of $OD-CLF$) but results in poor counting accuracy as the YOLO network in itself is trained exclusively for location and does not incorporate counts.

Both networks are trained end to end utilizing the same data sets, using the training and test sets as depicted in table \ref{table:3}.
Prior to training the model for $IC$, we first set the loss function hyper-parameters using data from the training set (See Section \ref{sec:ic}). For all three data sets, we use ADAM optimizer with learning rate $10^{-4}$ and exponential decay of $5\times10^{-4}$, and exit training when the performance on the validation set begins to drop. For the \textbf{Coral} and \textbf{Jackson} data sets 10 epochs of training were required to achieve optimal performance. 
The more complex \textbf{Detrac} dataset with three classes requires 20 epochs to converge. 
We train the model for $OD$ techniques using stochastic gradient descent with a exponential weight decay of $5\times10^{-4}$ and momentum of $0.9$. However, we chose to use a smaller learning rate of $10^{-4}$ due to unstable gradients at the added branch. We train \textbf{Coral} and \textbf{Jackson} for 4 epochs and the \textbf{Detrac} data set for 6 epochs. The hyper-parameters of the loss function were set manually based on the data from the training set (see Section \ref{sec:od}). For $OD$ techniques we threshold the grid cell to determine the presence of an object using a threshold of 0.2.

We implement our models with the PyTorch framework \cite{paszke2017automatic} and perform all experiments on a single Nvidia Titan XP GPU using an HP desktop with an Intel Xeon Processor E5-2650 v4, and 64GB of memory.

\subsection{Filter Accuracy}

\begin{figure}
\includegraphics[width=0.48\textwidth, trim={0cm 0cm 0cm 0cm}, clip]{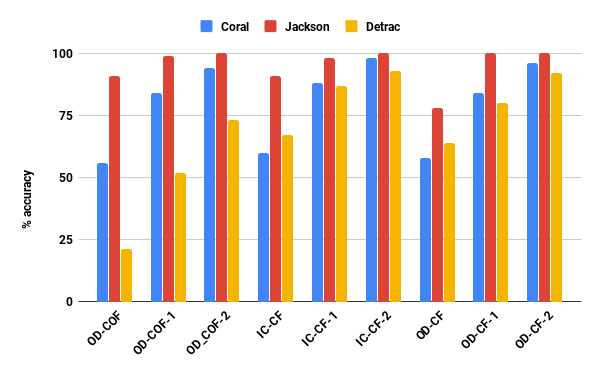}
\caption{\label{fig:acc-cf} Accuracy of object count filters on the three data sets }
\end{figure}

In this set of experiments we quantify the accuracy of the various proposed filters. Figure \ref{fig:acc-cf} presents the results of the accuracy of the $OD-COF$, $IC-CF$ and $OD-CF$ filters for the three data sets. For a frame $f$ let $c_f$ be the number of objects in $f$ and $\hat{c_f}$ be the filter estimate. Accuracy is defined as the fraction of frames for which $\hat{c_f}=c_f$.  We also present two variants for each filter annotated with postfix $1$ and $2$ to assess approximate counts. Filters $OD-COF-1$ (correspondingly $IC-CF-1$, $OD-CF-1$) quantify the fraction of frames $f$ for which $c_f-1 \leq \hat{c_f} \leq c_f+1$.
Similarly $OD-CF-2$ (correspondingly $IC-CF-2$, $OD-CF-2$) quantify the fraction of frames for which $c_f-2 \leq \hat{c_f} \leq c_f+2$.

It is evident that for the {\em Coral} data set all three techniques perform the same when estimating the exact count of objects in frames. Accuracy increases very quickly for all techniques for $CF-1$ and $CF-2$ filters. All techniques appear to equally underestimate and overestimate the true count of objects. In the case of this data set the average number of objects per frame is 8.5 and
the variance 5.1.  The situation is similar in the case of the {\em Jackson} data set; even though the data set has two classes of objects
the average number of objects per frame as well as the variance across frames is low and all three estimation techniques present comparable results.
In the case of {\em Detrac} the situation is a bit different. The average number of objects per frame is higher as well as the variance. Moreover the data set has three classes and the most popular class (car) has high variance in itself. In this case $OD-COF$ performs poorly; evidently utilizing the convolution features only for count estimation is ineffective as the number of objects per frame increases. The other two techniques are competitive and become much better for $CF-1$ and $CF-2$ filters (approximate counts). The $IC$ techniques perform slightly better than the $OD$ techniques for count estimation. Since the network for $IC$ is pre-trained on ImageNet to perform classification, the feature maps seem to
provide dense and discriminate context of the objects which appears more suited for transfer learning tasks such as counting. 
\begin{figure*}[!htb]
\minipage{.33\textwidth}
\includegraphics[width=\linewidth]{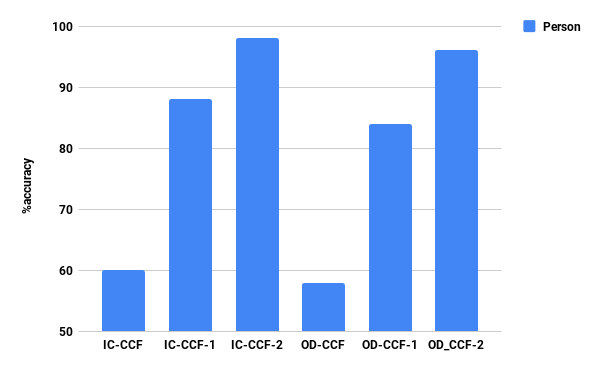}
\caption{Coral Dataset}
\label{fig:ccf1}
\endminipage\hfill
\minipage{.33\textwidth}
\includegraphics[width=\linewidth]{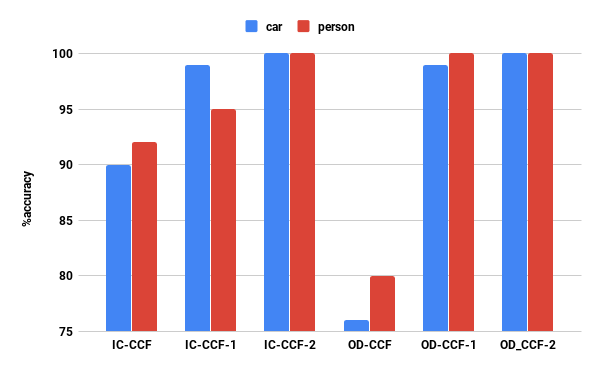}
\caption{Jackson Dataset}
\label{fig:ccf2}
\endminipage\hfill
\minipage{.33\textwidth}%
\includegraphics[width=\linewidth]{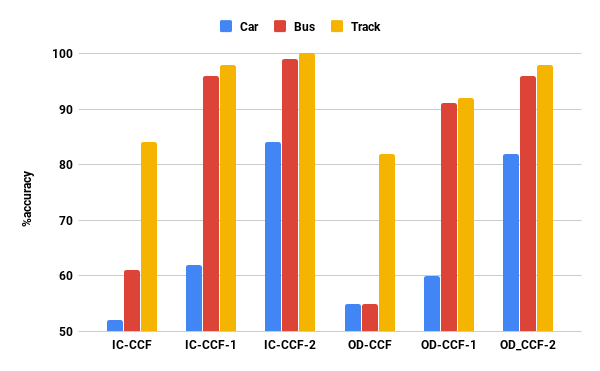}
\caption{Detrac Dataset}
\label{fig:ccf3}
\endminipage
\caption{$CCF$ performance across data sets for varying number of classes}
\label{fig:ccf}
\end{figure*}
This is evident in Figure \ref{fig:ccf} which presents the count estimates per class for the three data sets. For {\em Coral} (Figure \ref{fig:ccf1}) we observe that the two techniques perform almost equally. In the case of the {\em Jackson} data set (Figure \ref{fig:ccf2}) $IC$ techniques perform better when considering exact counts, but both techniques present similar accuracy for approximate counts (within count one and two). Finally the same trend is confirmed also on the {\em Detrac} data set (Figure \ref{fig:ccf3}). It is worth noticing that the techniques present higher accuracy for classes that are less popular. Even though the less popular classes in the video stream are represented by less training examples, less popular objects in frames have typically low counts in each frame. They thus constitute an easier estimation problem.
In summary $COF$ techniques do not appear competitive for cases with larger number of objects per frame. Moreover  $IC$ and $OD$, $CCF$ filters demonstrate comparable accuracy, with $IC-CCF$ filters having a slight advantage in exact counting.

\begin{figure*}
\begin{minipage}{.34\textwidth}
\includegraphics[width=\linewidth]{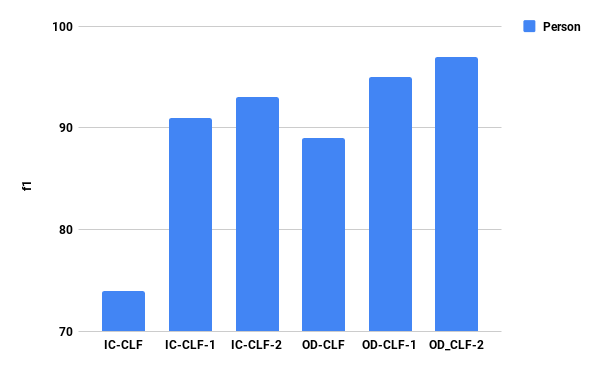}
\caption{Coral Dataset}
\label{fig:clf1}
\end{minipage}
\begin{minipage}{.34\textwidth}
\includegraphics[width=\linewidth]{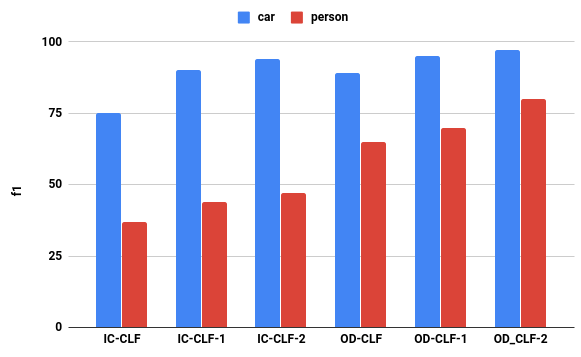}
\caption{Jackson Dataset}
\label{fig:clf2}
\end{minipage}
\begin{minipage}{.34\textwidth}
\includegraphics[width=\linewidth]{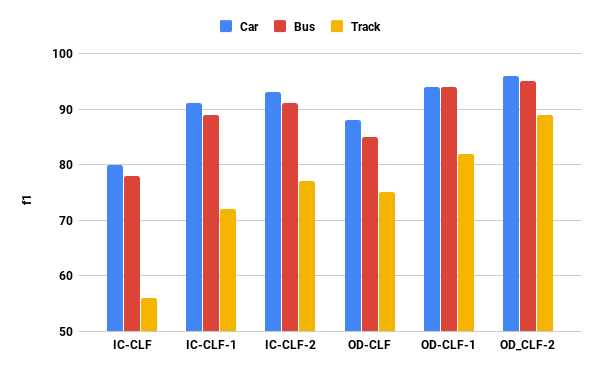}
\caption{Detrac Dataset}
\label{fig:clf3}
\end{minipage}
\caption{$CLF$ performance across data sets for varying number of classes.}
\label{fig:clf}
\end{figure*}

Figure \ref{fig:clf} presents the results for the case of $CLF$ filters. We aim to explore how well the two techniques are able to estimate object locations. 
Figure \ref{fig:clf1} presents the results for the case of the {\em Coral} data set. The $y$-axis measures the $f1$ score \footnote{Defined as a  function of true positives (tp: objects identified correctly in the validation set), false positives (fp: objects identified erroneously in the validation set) and false negatives (fn: objects failed to identify in the validation set) compared to ground truth. In particular $f1=\frac{2\times p \times r}{p+r}$, where $p=\frac{tp}{tp+fp}$ and $r=\frac{tp}{tp+fn}$} of the estimation. Each estimation takes place on a $56 \times 56$ grid in which we predict the presence of objects of a specific class. For each prediction we quantify the true positives, false positives and false negatives and compute precision and recall (and $f1$ measure) of object detection over all frames. $OD$ techniques demonstrate high accuracy localizing the exact grid cell that objects are located. As in the case of count estimation we also present two variants of the filter annotated with postfix $1$ and $2$. Filters $IC-CLF-1$ and $OD-CLF-1$ assess the localization of the grid cell prediction as correct if an object of the same class as the predicted one, lies in a cell at Manhattan distance one from the predicted cell (any of the four adjacent grid cells of the predicted one). Similarly for $IC-CLF-2$ and $OD-CLF-2$ filters, we expand the grid to Manhattan distance two around the predicted cell. The intuition is that spatial constraints are still preserved if the prediction is "close" (in cell distance) to the actual location of the object. We will evaluate the effectiveness and accuracy of these filters in sample queries. Overall $OD$ localization techniques perform much better as they are optimized for object location prediction. The network for $OD$ filters is an object detection network and thus the corresponding feature maps provide more details regarding the spatial and location features of the objects in the image.

For the {\em Jackson} data set (Figure \ref{fig:clf2}) as well as {\em Detrac} (Figure \ref{fig:clf3}) which have two and three classes respectively, $OD$ techniques demonstrate superior performance. We also observe that less popular classes (as is the case of person in {\em Jackson} and Truck and Bus in {\em Detrac}) have lower $f1$ score and this is an artifact of training; real videos contain unbalanced (in relative frequency of occurrence) object classes; less popular classes in the video stream present less training examples in the network and as a result the accuracy of location estimation is lower (unlike counts, predicting location is much harder).

In summary $OD$ techniques provide superior accuracy for localization and seem highly suitable for evaluating spatial constraints among objects in frames. At the same time they remain competitive for estimating object counts. 

We also conducted an experiment comparing $OD$ techniques to identify a constraint between two object classes (car left of a bus), against a data set that is manually annotated to recognize the constraint. Our results indicate that the proposed $OD$ filtering approach achieves superior accuracy, at 99\% against a manually annotated data set. However the additional benefit of our approach is {\em generality} as we do not require training for any possible class of object and any possible constraint among the objects.

\subsection{Query Results}
We now present an evaluation of the effectiveness of these filters in a query processing setting. We consider the following queries: For {\em Coral} data set: identify all frames with two people ($q_1$) as well as identify all frames with two people in the lower left quadrant of the frame ($q_2$). For {\em Jackson} data set, identify all frames with exactly one car and exactly one person ($q_3$), identify all frames with at least one car and at least one person ($q_4$), identify all frames with exactly one car and exactly one person and the car left of the person ($q_5$). For {\em Detrac}, identify the frames with exactly one car and exactly one bus ($q_6$) and identify the frames with exactly one car and exactly one bus and the car to the left of the bus ($q_7$). 

For these queries we progressively enable all applicable filters and for the frames that pass each filter we evaluate the entire frame with an object detector (Mask R-CNN) for the final evaluation. For the queries that involve only counts we assess accuracy as the fraction of frames that are correctly identified by our filters over the number of frames in which the query predicates are true (ground truth). For the queries that involve spatial constraints we report the $f1$ measure for each query. For count estimation and class count estimation we deploy $OD-CCF$ filters and for evaluating spatial constraints we deploy $OD-CLF$ filters. We also evaluate each query in a brute force manner annotating all frames with Mask R-CNN to determine the true frames in the query results. To run {\em Coral} through Mask R-CNN requires 5.2 hours, {\em Jackson} 3.89 hours and {\em Dectrac} 7.14 hours. Table \ref{table:4} presents the query results. In the figure we present the {\em most selective} filter combinations that yield 100\% accuracy (with the exception of query $q_7$ in which the resulting accuracy is 93\%). 
\begin{footnotesize}
\begin{table}

\begin{center}
\begin{tabular}{||c|c|c|c|c|c|c|c||}
\hline
Filter & $q_1$ & $q_2$ & $q_3$ & $q_4$ & $q_5$ & $q_6$ & $q_7$ \\ \hline
\tiny{OD-CCF} & - & - & 87.4 & 122.6 & - & - & -\\ \hline
\tiny{OD-CCF-1} & 909.4 & - & - & - & - & 367.6 & -\\ \hline
\tiny{OD-CCF-1/OF-CLF} & - & 427 & - & - & - & - & -\\ \hline
\tiny{OD-CCF/OD-CLF-1} & - & - & - & - & 67.6 & - & -\\ \hline
\tiny{OD-CCF-1/OD-CLF-2} & - & - & - & - & - & - & 293.4* \\ \hline
\end{tabular}

\caption{\label{table:4} Execution times (sec) and filter combinations to achieve 100\% accuracy ($q_7$ accuracy is 93\%).}

\end{center}
\end{table}
\end{footnotesize}

As is evident from the table, the performance advantage offered by the filters proposed are very large. In most of the cases, without loss in accuracy the filters offer orders of magnitude performance improvements, enabling running complex queries on video streams much faster than it was possible before.

\subsection{Aggregates}
We now turn to evaluate correlated aggregates and their applicability to reduce the variance for aggregate estimation of queries involving multiple objects as well as spatial constraints among them. We focus on five sample queries {\em in all 3 data sets}, namely in the {\em Jackson} data set, estimate the number of frames having a car on the lower right quadrant ($a_1$) as well as a query estimating the number of frames with a car on the left of a person ($a_2$), for {\em Detrac} estimate the number of frames with three objects and a car in the lower left quadrant and a bus in the upper left ($a_3$) and estimate the number of frames with a car left of a bus ($a_4$). Finally for {\em Coral} estimate the number of frames with three people out of which at least two people are in the lower left quadrant ($a_5$). The results are depicted in Table \ref{table:5}. We execute the query sampling randomly from the stream; each query is executed one hundred times and we report averages. The straightforward way to do the estimation is to evaluate each sampled frame with Mask R-CNN and derive a statistical estimate of the results.  Applying correlated aggregates to enable the proposed filters to derive an estimate of the correlated aggregate and then apply Mask R-CNN. In the final estimation we utilize the correlated aggregate aiming to reduce the variance of the estimate (as per Section \ref{sec:aggr}). We report the time required per frame for our filters (correlated variables) followed by and including Mask R-CNN reporting on the corresponding reduction in variance for each query estimate. The time to run Mask R-CNN in each frame is 200ms. It is evident that variance is reduced substantially with a small increase in the processing time per frame.
\begin{footnotesize}
\begin{table}
\begin{center}
\begin{tabular}{||c|c|c||}
\hline
Query & Filter + Mask RCNN & Variance Reduction \\ \hline
$a_1$ & 201.6ms & 48 \\ \hline
$a_2$ & 201.6ms & 12 \\ \hline
$a_3$ & 202.2ms & 38 \\ \hline
$a_4$ & 201.6ms & 230 \\ \hline
$a_5$ & 202.2ms & 89 \\ \hline
\end{tabular}

\caption{\label{table:5} Sample estimation queries, time per frame to enable applicable filters and Mask R-CNN execution and the resulting reduction in variance.}
\end{center}
\end{table}
\end{footnotesize}
\section{Related Work}
\label{sec:related}

Recently there has been increased interest in the application of Deep Learning techniques in data management \cite{DBLP:conf/sigmod/KraskaBCDP18, DBLP:conf/cidr/MarcusP19,DBLP:journals/corr/abs-1903-09999,DBLP:conf/icde/ChengK19,DBLP:journals/corr/abs-1903-10000}.
NoScope \cite{Kang:2017:NON:3137628.3137664} initiates work in surveillance video query processing. The authors address fast query processing on video streams focusing on frame classification queries, namely identify frames that contain certain classes of objects involved in the query. They train deep classifiers to recognize only specific objects, thus being able to utilize smaller and faster networks. They demonstrate that filtering for specific query objects can improve query processing rate significantly with a small loss in accuracy. In subsequent work \cite{blazeit,DBLP:conf/cidr/KangBZ19} the authors introduce a query language for expressing certain types of queries on video streams. They also discuss sampling based techniques inspired by approximate query processing to answer certain types of approximate aggregate queries on a video stream as well as use control variates techniques from the literature \cite{glasserman}, for a single variable to reduce variance of aggregates. Our work builds upon and extends these works by focusing on query processing taking into account spatial constraints between objects in a frame, {\em a problem not addressed thus far}, as well as demonstrating how to adapt control variates for multiple variables since in our problem multiple objects are involved in a query possibly with constraints among them. 
Lu et. al., \cite{Lu:2016:ORP:2987550.2987564} present Optasia, a system that accepts input from multiple cameras and applies difference video query processing algorithms from the literature to piece together a video query processing system. The emphasis of their work is on system aspects such as parallelism of query plans based on number of cameras and operation complexity as well as parallelism to allow multiple tasks to process the feed of a single camera.

Stream query processing is a well established area in the database community \cite{Koudas:2003:DSQ:1315451.1315572,DBLP:journals/fttcs/Muthukrishnan05}. Although the bulk of the work focused on numerical and categorical streams, numerous concepts invented apply in the streaming video domain \cite{DBLP:series/ads/DatarM07,DBLP:journals/pvldb/MankuM12,DBLP:journals/tkde/GuhaMMMO03,DBLP:conf/sigmod/AnanthanarayananBDGJQRRSV13,DBLP:conf/vldb/GolabO03,DBLP:conf/sigmod/LiMTPT05,DBLP:conf/edbt/ZhangGTS02,DBLP:conf/sigmod/SrivastavaKZO05,DBLP:journals/sigmod/LiMTPT05,DBLP:conf/icde/QinQZ06,DBLP:conf/icde/NagarajNRS08,DBLP:conf/icdt/Woodruff16}. In particular work on operator ordering \cite{Babcock:2003:COS:872757.872789,Lu:2018:AML:3183713.3183751} is highly relevant when considering multiple filters to reduce the number of frames processed. We foresee a resurgence of research interest in this areas taking the characteristics of video data into account. Spatial database management \cite{DBLP:books/aw/Samet90} is another well establish field in data management from which numerous concepts apply when modeling objects in an image and their relationships, in the case of streaming video query processing. In particular past work on topological relationships on spatial objects \cite{Papadias:1995:TRW:223784.223798} is readily applicable.

Approximate queries have been well studied in the database community \cite{Chaudhuri:2017:AQP:3035918.3056097,Park:2018:VUA:3183713.3196905,Rusu:2006:FRR:1142473.1142496,Chaudhuri:2007:OSS:1242524.1242526,DBLP:journals/vldb/0003KOSZ10,DBLP:journals/pvldb/PottiP15,DBLP:conf/sigmod/Kraska17,DBLP:journals/pvldb/GalakatosCZBK17,DBLP:conf/sigmod/PengZWP18}. Numerous results for different types of queries exist with varying degrees of accuracy guarantees. These results are readily applicable to different types of queries of interest in a streaming video query processing scenario. 

Recent results in the computer vision community have revolutionized object classification \cite{Krizhevsky:2017:ICD:3098997.3065386,DBLP:journals/corr/SimonyanZ14a,DBLP:conf/cvpr/SzegedyLJSRAEVR15,DBLP:conf/cvpr/HeZRS16} as well as object detection \cite{DBLP:conf/iccv/Girshick15,DBLP:conf/cvpr/GirshickDDM14,DBLP:journals/pami/RenHG017,DBLP:conf/iccv/HeGDG17,DBLP:conf/cvpr/RedmonDGF16,DBLP:conf/cvpr/RedmonF17,DBLP:journals/corr/abs-1804-02767}. Among object detection approaches the family of R-CNN \cite{DBLP:conf/iccv/Girshick15,DBLP:conf/cvpr/GirshickDDM14,DBLP:journals/pami/RenHG017,DBLP:conf/iccv/HeGDG17} papers achieves strong results, but the area is still under rapid improvement. Our proposed $OD$ techniques inspired by object detection utilize the YOLOv2 \cite{DBLP:conf/cvpr/RedmonDGF16,DBLP:conf/cvpr/RedmonF17,DBLP:journals/corr/abs-1804-02767} architecture, but can easily adapt any detection framework since all are convolutional with main differences in the way the actual objects are extracted (YOLOv2 uses ideas from R-CNN as well). Similarly our proposed $IC$ techniques utilizing classification are based on VGG19 \cite{DBLP:journals/corr/SimonyanZ14a} but can easily adapt any classification framework. Several recent surveys, summarize the results in the areas of object classification and detection \cite{DBLP:journals/cviu/GirshickKLMPVWY17}. The properties of deep features learned during training convolutional networks for localization have been utilized before \cite{DBLP:conf/wacv/BazzaniBAT16,DBLP:journals/pami/CinbisVS17,DBLP:conf/cvpr/OquabBLS14,DBLP:conf/cvpr/OquabBLS15,DBLP:conf/cvpr/ZhouKLOT16}. Here we utilize this observation for counting.
Density map estimation (number of people present per pixel in an image) is a problem central in crowd counting \cite{DBLP:journals/corr/SindagiP17}. The main emphasis has been on images containing hundreds or thousands of objects (e.g., people, animals, etc) with specific annotations (dot annotations). 
In contrast we focus on applications with small number of objects per frame, training networks of limited size with emphasis on performance, so the techniques can be easily applied along with standard training methods, for query evaluation. Moreover, we are also addressing the problem of counting per object class, which is not the focus on crowd counting approaches. Our motivation stems from query processing as opposed to counting crowds.\\
A system demonstration and a prototype system encompassing the techniques and concepts introduced in this paper is available elsewhere \cite{xarchakos}.

\section{Conclusions}
\label{sec:conc}

We have presented a series of filters to estimate the number of objects in a frame, the number of objects of specific classes in a frame as well as to assess an estimate of the spatial position of an object in a frame enabling us to reason about spatial constraints. These filters were evaluated for accuracy and we experimentally demonstrated using real video data sets that they attain good accuracy for counting and location estimation purposes. When applied in query scenarios over video streams, we demonstrated that they achieve dramatic speedups by several orders of magnitude. We also presented techniques to complement our video monitoring study, that reduce the variance of aggregate queries involving counting and spatial predicates.

This work opens numerous avenues for further study. Declarative query languages and query processors for video streams is largely an open research area. Studying query optimization issues in our framework is an important research direction. Study of additional query types involving spatial and temporal predicates is a natural extension. Finally extension of the filters for crowd counting and estimation scenarios is also necessary. 

\bibliographystyle{abbrv}
\bibliography{videoquery}

\end{document}